\documentclass{vgtc}

\usepackage{subcaption}
\usepackage{graphicx}
\usepackage{float}

 \renewcommand{\copyrighttext}{978-1-5386-8194-7/18/\$31.00
\copyright{}\,2018 IEEE}
\makeatletter
\def\@oddfoot{\hfil\sffamily\small\MakeUppercase{2018 IEEE Symposium on
Visualization for Cyber Security (VizSec)}\hfil}
\renewcommand{\ps@empty}{\renewcommand{\@oddfoot}{\hfil{\small\sffamily\copyrighttext}\hfil}}
\makeatother
\usepackage{microtype}                 
\PassOptionsToPackage{warn}{textcomp}  
\usepackage{textcomp}                  
\usepackage{mathptmx}                  
\usepackage{times}                     
\usepackage{cite}                      
\usepackage{pifont}

\usepackage{tikz}
\usepackage{standalone}
\usetikzlibrary{calc,patterns,decorations.pathmorphing,decorations.markings}
\usetikzlibrary{positioning}
\usetikzlibrary{shapes,decorations}
\usetikzlibrary{shapes.geometric, arrows, positioning}
\usepackage{pgfplots}
\pgfplotsset{compat=1.14}
\usepackage{cite}
\usepackage{xcolor}
\usepackage{todonotes}

\vgtccategory{Research}

\CCScatlist{
  \CCScatTwelve{Human-centered computing}{Visu\-al\-iza\-tion}{Visu\-al\-iza\-tion techniques}{Graph drawings};
  \CCScatTwelve{Human-centered computing}{Visu\-al\-iza\-tion}{Visualization systems and tools}{Visualization toolkits};
  \CCScatTwelve{Security and privacy}{Systems Security}{Vulnerability management}{};
  \CCScatTwelve{Security and privacy}{Security in hardware}{Embedded systems security}{}
}

\begin{document}

\title{Looking for a Black Cat in a Dark Room:\\ Security Visualization for Cyber-Physical System Design and Analysis}

\author{Georgios Bakirtzis\thanks{Electrical \& Computer Engineering, Virginia Commonwealth University, Richmond, VA, USA (email: bakirtzisg@ieee.org)}\\ %
\and Brandon J. Simon\thanks{Electrical \& Computer Engineering, Virginia Commonwealth University, Richmond, VA, USA (email: simonbj@vcu.edu)}\\ %
\and Cody H. Fleming\thanks{Systems Engineering and
Mechanical \& Aerospace Engineering, University of Virginia, Charlottesville, VA, USA (email: fleming@virginia.edu)}
\and Carl R. Elks\thanks{Electrical \& Computer Engineering, Virginia Commonwealth University, Richmond, VA, USA (email: crelks@vcu.edu)}%
}

\abstract{
Today, there is a plethora 
of software security tools employing visualizations that enable the creation 
of useful and effective interactive security analyst dashboards. 
Such dashboards can assist the analyst 
to understand the data at hand 
and, consequently, to conceive more targeted preemption 
and mitigation security strategies.
Despite the recent advances,
model-based security analysis is lacking tools
that employ effective dashboards---to
manage potential attack vectors, system components,
and requirements.
This problem is further exacerbated 
because model-based security analysis produces significantly larger result spaces
than security analysis applied
to realized systems---where platform specific information, software versions, 
and system element dependencies are known.
Therefore, there is a need 
to manage the analysis complexity
in model-based security
through better visualization techniques.
Towards that goal, we propose an interactive security analysis dashboard 
that provides different views largely centered around the system, its requirements, 
and its associated attack vector space.
This tool makes it possible 
to start analysis earlier in the system lifecycle.
We apply this tool in a significant area of engineering design---the design 
of cyber-physical systems---where security violations can lead to safety hazards.
}

\firstsection{Introduction}
\label{sec:intro}
\maketitle

Security visualizations have changed the way we view, organize,
and respond to system violations.
The study of effective visualizations
for security has led
to better mitigation strategies applied
in real time~\cite{marty:2009}.
Nevertheless, there are two areas where visualization
has made little progress:
assessing the security posture 
of competing design patterns early
in the system's lifecycle
and effective visualization
for large amounts of evidence generated 
at the early design phase.

This issue was not as critical when systems took years
to develop and deploy---where designs went
through several rounds of testing, specifications where clearly defined,
and a main architect took responsibility 
for the development of the system.
Recently, however, there has been a shift 
to deploying system's designed using commercial-off-the-shelf (\textsc{cots}) hardware
on top of open-source software,
which has the \emph{potential} of reducing technical debt
and financial cost.
On top of this, non-safety-critical and safety-critical systems alike are increasingly connected 
to the internet.
(See, for example, the security 
and privacy issues surrounding the industrial internet 
of things~\cite{sadeghi:2015}.)

For this reason, a significant challenge is the secure design
and deployment of cyber-physical systems (\textsc{cps}), 
where security violations can lead to unsafe physical behavior~\cite{moreno:2018}.
Effective visualizations 
in this area would achieve a higher degree 
of operational assurance with respect
to security 
in applications where security violations could lead to safety hazards, such as 
medical systems~\cite{alemzadeh:2013}, aviation~\cite{boeing},
automotive~\cite{checkoway:2011,computest:2018,keenlab:2018}
and electric power~\cite{kshetri:2017}, 
to name a few.

Furthermore, the use of systematic
and standardized analysis through effective visualization
in the \textsc{cps} domain provides a common language
between security professionals and system designers, which is currently limited. 
The lack of such language in the security domain
can have detrimental effects, including miscommunication
of security requirements to system designers, security applied for the sake
of security (not based on real operational needs),
and security obstructing system requirements.

Additionally, a major problem in applying security
as a lifecycle practice is that the amount
of data generated at the design phase is significantly larger
than that used for the analysis of realized systems.
This is because, the inherent incompleteness of information
at the design phase produces a large amount of applicable attack vectors.
Lacking information includes but is not limited to
the specific versions of software,
the security patches that will be applied to the system,
and the potential insecurity caused by the coupled system.

Moreover, security data is recorded
to be parsed and used by security professionals. 
Security analysts are expected
to be looking for specific vulnerabilities 
and attack patterns to report in a standardized fashion---making
implicit assumptions 
and using expert knowledge from previous experiences.
The current challenge is
to allow system designers to use security data.
This is one way that system security can be exercised 
as a lifecycle practice;
especially earlier in the lifecycle 
where decision effectiveness is highest~\cite{saravi:2008, frola:1984, strafaci:2008}.


\begin{figure*}
\begin{subfigure}[b]{1.0\textwidth}
   \includegraphics[width=1.0\textwidth]{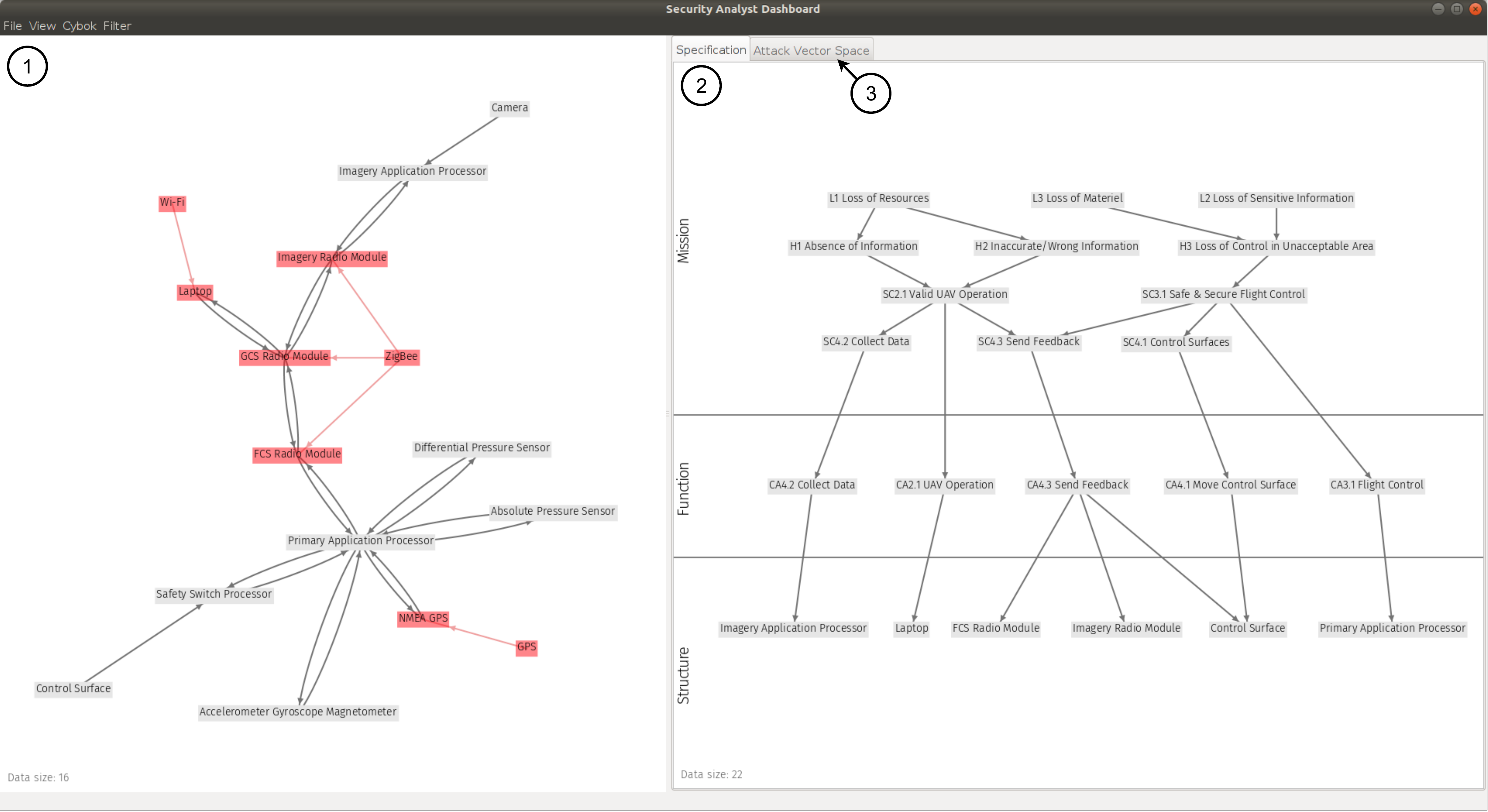}
   \caption{}
   \label{fig:dashboard1} 
\end{subfigure}

\begin{subfigure}[b]{1.0\textwidth}
   \includegraphics[width=1\linewidth]{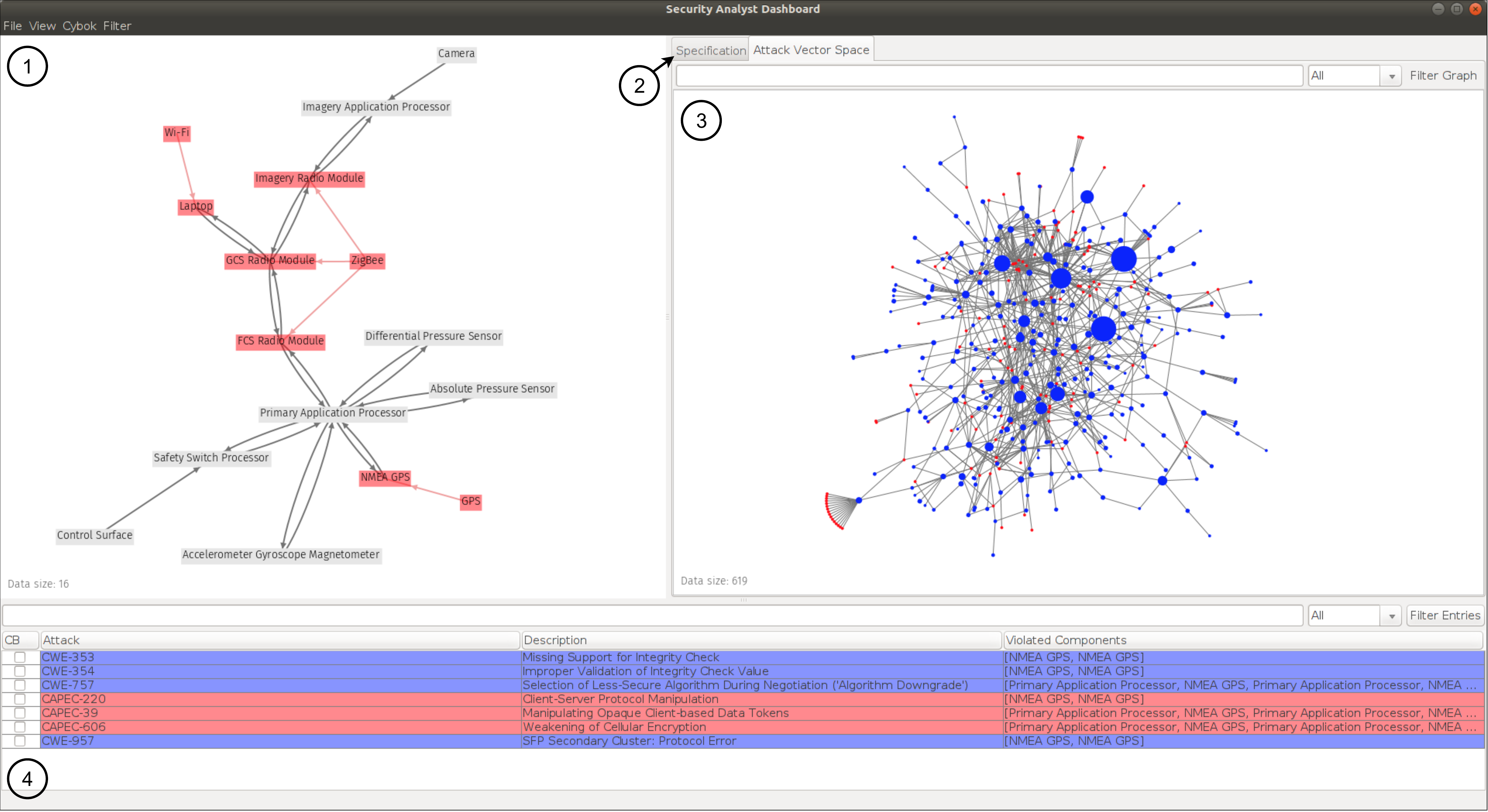}
   \caption{}
	\label{fig:dashboard2}
\end{subfigure}
\caption[Two main views of the security dashboard]{Screenshots of the user interface of the proposed security dashboard. The dashboard supports diverse information for better informing system designers and security analysts alike of the goals of the system (specification), the potential attack vectors that can violate the goals of the system, and the system attack surface projected over the system topology.}
\label{fig:dashboard}
\end{figure*}

To combat these challenges, we implement a security analyst dashboard
that can be used preemptively from the early stages
of a system's lifecycle to deployment and operation,
where resilience or hardening defenses might be added. 

\textbf{Contributions.} \quad The contributions of this work are:
\begin{itemize}
\item We develop an open-source security analyst dashboard that supports system designers and security analysts  alike;
in turn, this provides a common language between the two.
\item We provide important functionality to deal with the large number of data intrinsically produced when applying security analysis in the absence of a realized system.
\item We present system and operational information lacking from current security analysis and visualization tools that define the mission of the system
and, consequently, allow tracing potential security violations
to degradation of mission-level requirements.
\end{itemize}
\vspace{1cm}

\section{Security Analyst Dashboard}
\subsection{Domain problem}

In the design 
of complex systems in general---but specifically in the area of \textsc{cps} design 
and analysis---it is important
to consolidate information
from a diverse set
of stakeholders.
These include but are not limited
to the operations staff, end users,
and subject domain experts.
System designers and security analysts need
to consider all this information 
when proposing a possible design solution.
Furthermore, it is becoming increasingly evident
that system designers and security analysts
require a common language,
such that security considerations are applied proactively
during system design 
and throughout the system's lifecycle.

By definition, the amount
of information to consider at the design phase
is significantly larger
and more complex to navigate 
compared to the security analysis
of a realized system.
This problem is further extended in \textsc{cps}
by the need to avoid hazardous and unsafe behavior.
In addition to potential system violations; that is, exploitation of system resources, 
system designers and security analysts need
to be aware of the requirements,
for example, unacceptable losses, potential hazards, unsafe control actions,
and admissible functions
of the system. 


To overcome these challenges we propose 
to collectively analyze both safety 
and traditional attack vector artifacts
based on a system topology model,
which allows the analyst
to provide a thorough and complete security report.
To achieve this effectively, the dashboard 
presents and allows for navigating the large number 
of informational artifacts generated
at the design phase through several graphs. 
The two primary means
of analysis conducted using the dashboard are:

\begin{itemize}
\item \textbf{Systems-theoretic analysis} (top-to-bottom) \quad
Security violations are emergent properties of the system.
Emergent properties stem from the coupling
of subsystems and cannot be investigated
by examining the subsystems individually.
This means that the assessment
of individual elements of the whole system
is insufficient to assure safe and secure behavior.
Furthermore, security needs to be exercised
to the extent necessary based 
on the system's expected service~\cite{bellovin:2015}. 
Therefore, this analysis relies
on data collected 
through a structured elicitation process~\cite{carter:2018a}
and is captured in Systems Modeling Language (SysML)~\cite{carter:2018b};
a familiar and often used modeling language to Systems Engineers.
The model is then transformed automatically
to a graph using GraphML~\cite{bakirtzis:2018b,brandes:2013}.
The purpose of using a GraphML metamodel is to allow agnosticism
towards modeling tool or language.
The resulting artifacts allow system
and security analysts
to reason about defenses
and potential violations
in a system's design without requiring a prototype realization
for testing potential hypothesis on.
This, in turn, allows for quick comparisons
of potential designs with respect to their security posture.
While this information is a natural consequence
of good systems engineering, it is often omitted
from security analysis.
Consequently, security is applied 
in a reactive \textit{bolt-on} fashion 
instead of being applied proactively---at the early stages 
of system design---and reapplied 
and assessed thereafter---up to deployment and operation.
By consolidating this information
in a security analyst dashboard
we present the systems
and its operational goals as equally important
to traditional attack vector analysis.

\item \textbf{Attack vector analysis} (bottom-to-top) \quad Following the top-to-bottom analysis there is a crucial stage,
where traditional vulnerability analysis
is applied in a model-based setting.
By exploring and filtering the attack vector space it is possible
to construct a defensible evidential trace
of potential violations in the system.
This is possible by adding extra design information
in the system model that can map
to historically recorded attack vectors~\cite{bakirtzis:2018a}.
To do this automatically we use techniques
from computational linguistics 
in conjunction with the exported GraphML models~\cite{bakirtzis:2018c}.
This analysis complements the top-to-bottom analysis
with real attack vectors, \textit{evidence}, in the absence
of a realized system.
By doing so, we are able to reduce the criticality 
of elements that might seem crucial using only systems-theoretic means
but are unlikely to be successfully violated.
We note, that this approach could be extended
to private repositories of companies
or agencies to extend the amount
of known attack vectors.

\end{itemize}

\subsection{System \& specification models}

As systems become more complex, how does one know what security needs exist (e.g., what should or should not be secured) if one does not know what the system needs to do (or must not do)?
Visualization can help answer this question, but this begs a further question: what must be visualized? One answer comes from the principles of systems engineering, which attempts to design and manage complex systems through the development of system requirements, understanding how the system should behave functionally, and creating an interacting set of components (i.e., an architecture) that achieves these behaviors. 
In the context of security, one must define what the system \textit{should not do}, in addition to requirements capturing what the system \textit{should} do.

Taken together, these artifacts---(1) system requirements, (2) functional behavior, and (3) architecture---result in a specification. This specification then forms the intellectual basis for visualization; that is, what should be visualized and why. For example, an analyst investigating a particular component might be able to visualize the components it interacts with (via the architecture), the kinds of behaviors these components give rise to (via the functional diagrams), and ultimately the expected service that the component is critical to (via the requirements).

This is especially important in assessing the security posture
of \textsc{cps}, because of the tight integration
of digital control with the physical environment.
The specification should capture what impact
can result from the violation
of digital components
and how that impact is reflected in the physical world.

Therefore, the model includes all important attributes
that are necessary for a holistic consequential security analysis.
That is, it includes the specification, the admissible behaviors,
and the system topology that implements the admissible behaviors.
To use the model with the dashboard we transform the SysML definitions
of the requirement diagrams 
and internal block diagrams
to graphs. 
The system topology graph is transformed 
to a directed graph $\Sigma = (V, E)$, 
where the vertices, $V(\Sigma)$, are the assets of the embedded \textsc{cps} 
and the edges, $E$, a dependence relationship between system assets (Figure~\ref{fig:dashboard} \ding{172}).
The specification is transformed 
to a directed hierarchical constrained layout graph, $S = (V, E)$, 
such that the different levels are independent
and $V(S)$ the unacceptable losses, potential hazards, safety constraints,
and critical components of the system topology graph, $\Sigma$ (Figure~\ref{fig:dashboard1} \ding{173}).

Normally, such a specification is captured
in tables. 
However, to fully leverage the specification
the analyst should be able to immediately see the relationships
that exist between different types
of information.
This is especially the case when the analyst wants 
to examine the impact
of either a specific attack vector applicable
to critical system elements
or when assuming that element will be violated, 
based on the analyst's previous experience.

\begin{figure}[!t]
\centering
\includegraphics[width=1.0\linewidth]{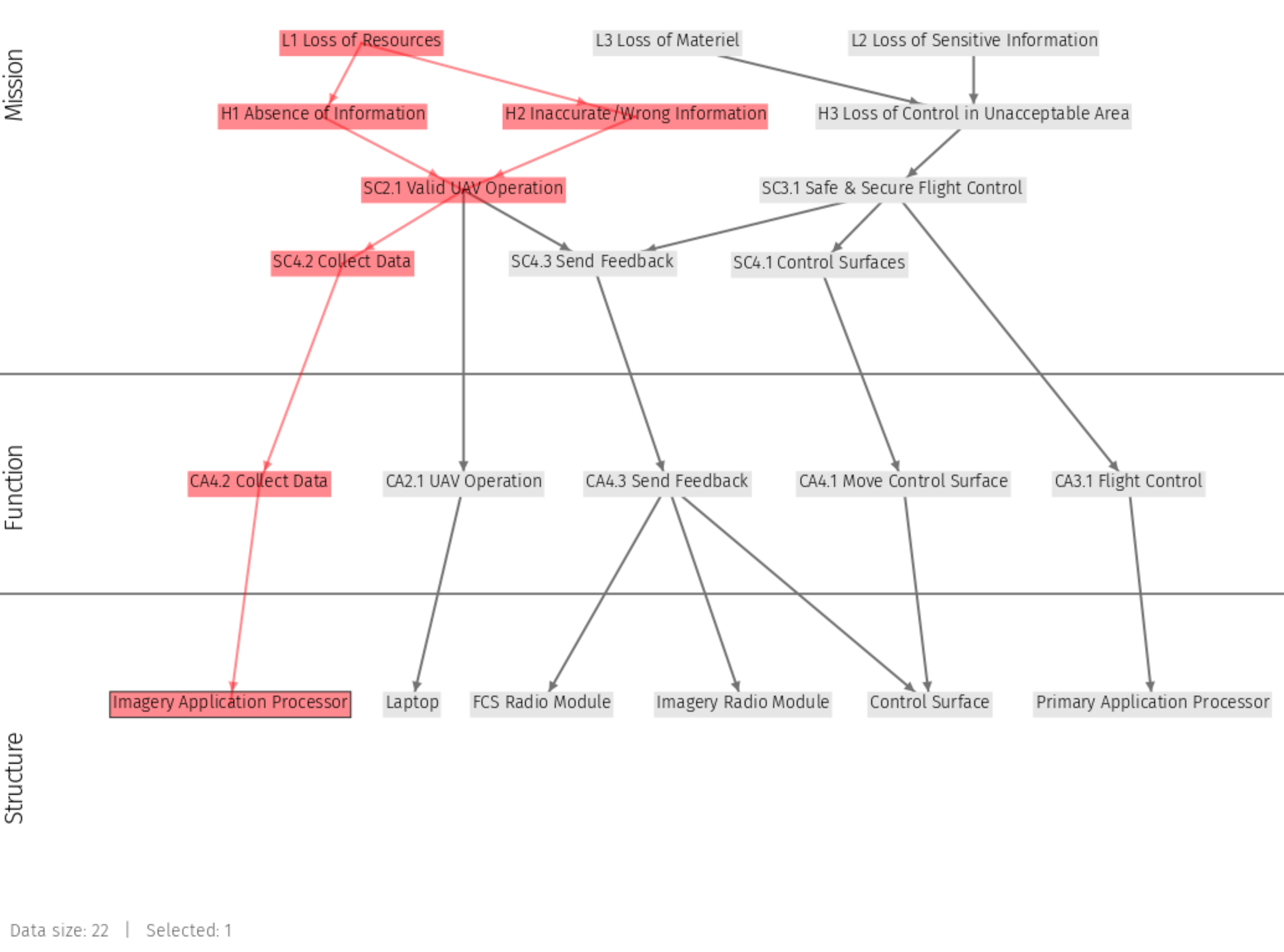}
\caption{Screenshot of the dashboard showing the completed mission specification, including a chain of violations in the event that the \textit{Imagery Application Processor} is violated through an attack vector.}
\label{fig:paths}
\end{figure}

\begin{figure*}[!t]
\centering
\includegraphics[width=1.0\linewidth]{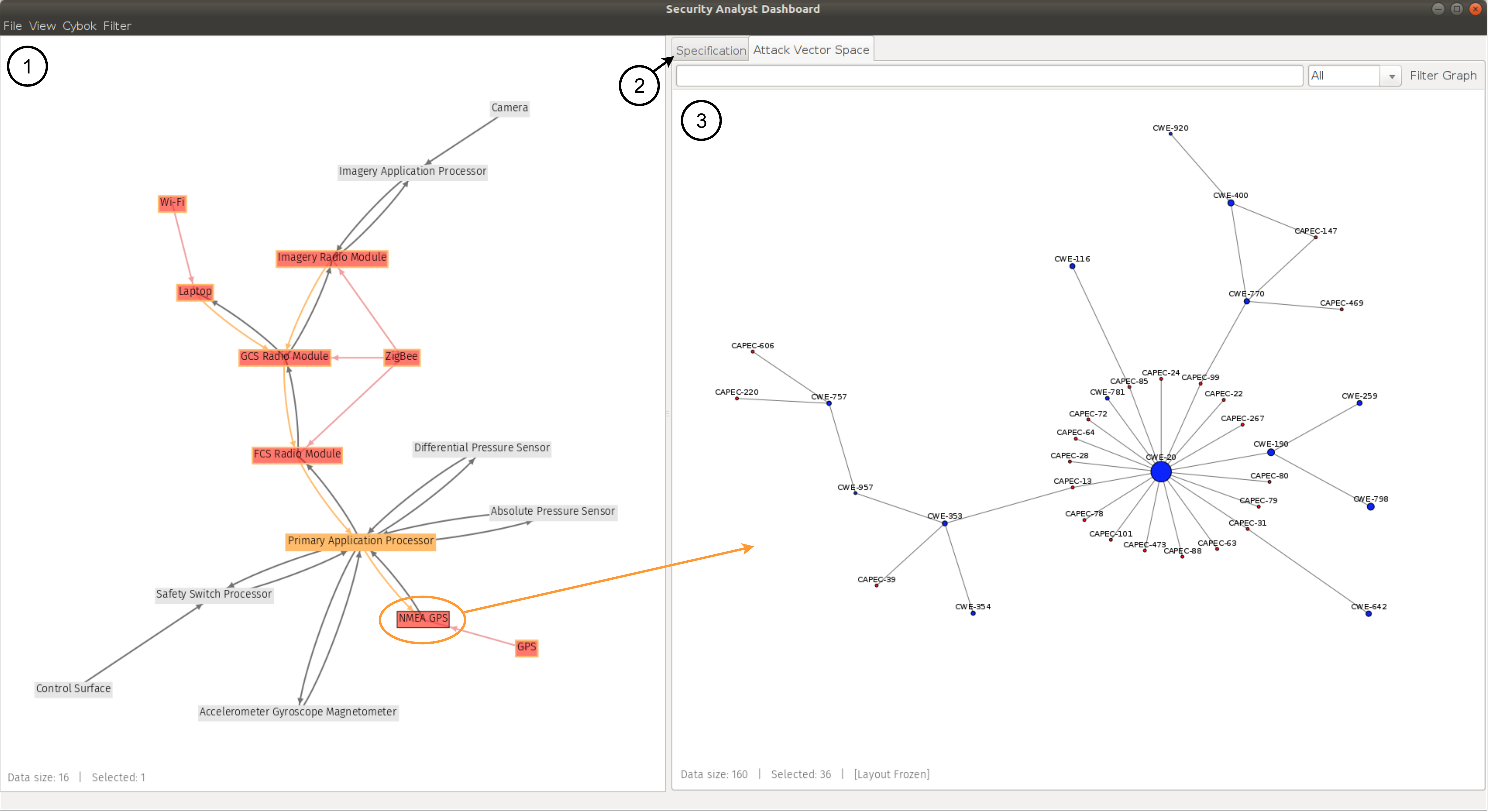}
\caption{Screenshot of the dashboard showing exploit chains in the system topology graph (\ding{172}) and filtering of the attack vector space per component; in this case NMEA GPS (\ding{174}), which allows the user to produce reportable evidence of potential violation in the system topology.}
\label{fig:expoitChains}
\end{figure*}

\subsection{Attack vector datasets}

In addition to the construction of models
for describing the requirements, functions,
and topology of the system we use open datasets to describe attack patterns, weaknesses,
and vulnerabilities.
These datasets have the additional benefit
of having a hierarchical taxonomy.
Specifically, the datasets used by the security dashboard are \textsc{mitre} Common Attack Pattern Enumeration and Classification (\textsc{capec})~\cite{CAPEC}, \textsc{mitre} Common Weakness Enumeration (\textsc{cwe})~\cite{CWE}, and \textsc{nist} National Vulnerability Database (\textsc{nvd})~\cite{NVD}, 
which recently took control
of \textsc{mitre} Common Vulnerabilities and Exposures (\textsc{cve})~\cite{CVE}.

In general, attack patterns associate with weaknesses
and vulnerabilities can abstract to weakness classes.
Given this intrinsic relationship between the databases
it is possible to represent the attack vector space
as a combined undirected attack vector graph $AV = (V, E)$, 
where $V(AV)$ the total collection of entries
of the datasets and $E$ the intra-related edges within each dataset
and inter-related edges between dataset entries (Figure~\ref{fig:dashboard2} \ding{174}).
This provides an intuitive basis
for exploring, filtering, and categorizing entries visually.

A significant concern 
in visualizing these datasets as a graph  
is the number of entries in each dataset.
To date, there are approximately one hundred 
and fourteen thousand entries in total, 
most of which are from \textsc{nvd}.
Therefore, even though \textsc{nvd} is extensive
and required to be thorough with the analysis
at the fidelity of the model, most of them can be abstracted
to classes of weaknesses as defined in \textsc{cwe}.
By utilizing this abstraction the number of total entries
is reduced to the size of the \textsc{capec} and \textsc{cwe}
datasets (around twelve hundred all together).
Therefore, at the worse-case scenario the starting attack vector graph
the analyst has to consider is significantly smaller
but equivalently descriptive
without including all \textsc{cve} entries.

Whilst a graph of this size can become overly complex
and difficult to navigate, it is a natural design choice
for the attack vector space associated with a system.
This is because a graph representation 
precisely captures the \emph{neighborhoods}
of attack vectors related to the system.
To manage this complexity we add intuitive interactivity
and filtering functionality such that the analyst can quickly
narrow down the results to what is relevant (Section~\ref{sec:interactivity}).

Additionally, by using this data, the analyst supplements a speculative \textit{what-if} analysis
with specific pieces of evidence
that such violations are possible in the system.
This information can then be used
to inform the rest of the stakeholders
to decide further applicable defensive requirements.

\subsection{Visualization \& interaction design}
\label{sec:interactivity}
Compared to real-time monitoring visualization tools, dashboards used
for in-depth analysis require a sophisticated set 
of interactivity functions 
to facilitate effective exploration
of diverse types of data~\cite{jacobs:2014}.
For example, by examining the system topology the analyst is also informed
about either the specification or the attack vector space.
This makes the individual data dynamic with respect to each other
and not just static complementary views.

Towards that goal, the main frame
of the security analyst dashboard organizes the information
in three main panes:

\begin{itemize}
\item \textbf{System topology ($\Sigma$).} \quad 
The system topology graph is a visualization of the design under analysis.
It also contains security specific information, including the attack surface of the system~\cite{manadhata:2010};
that is, the elements that a potential intruder can enter from based on found recorded historic attacks
at the entry point of the element (indicated by red vertices). 
Additionally, given a specific element to violate,
we draw the path of the potential exploit chains that can lead to its violation; that is, all paths from any element of the attack surface
to the chosen element,
where attack vectors have been found for the full path---both for the vertices and edges in that path---which is indicated by yellow vertices and edges (Figure~\ref{fig:expoitChains} \ding{172}).
Using this visualization the analyst can gather insights about the security posture
of the system without having to investigate individual attack vectors.
Simply by using visual cues it is possible to gain quick ideas about where
and how to add defenses or apply resilience techniques.

\item \textbf{Specification ($S$).} \quad         
The specification graph provides a view to the expected service from the position
of its requirements; that is, reiterated from before, the unacceptable losses, the potential hazards the system might have
during deployment and operation, and the safety constraints.
which define its overall mission, its control actions, the functionality necessary to achieve its mission, 
and finally, the critical system elements that in the case of violation will lead to the violation of other
necessary system functions.

The layout of the graph is custom; it defines constraint intervals based on the category of the vertices:
(1) the mission-level requirements, (2) the functional requirements, 
and (3) the elements of the system topology structure that are part of the specification.
By doing so, it shows the specification as a natural visual hierarchy, with the mission at the top, the function in the middle,
and the structure at the bottom.
An important action available in the specification graph is seeing
at any level what violations happen above or below it.
For example, an analyst can click on a structural element 
and see what functions and mission requirements are also violated, which allows him 
to understand the degree of mission degradation that occurs in that instance (Figure~\ref{fig:paths}).

\item \textbf{Attack vector space ($AV$).} \quad 
The attack vector graph shows all the attack vectors 
that could potentially violate any given component in the system topology.
Moreover, this graph visualizes the inter-related and intra-related connections
that are intrinsic between the vertices because of the hierarchical nature of the datasets;
that is, \textsc{capec}, indicated in red, and \textsc{cwe}, indicated in blue.
The entries of \textsc{cve} are abstracted
to their corresponding \textsc{cwe} classification.
This achieves a significantly lower number of vertices to visualize
and explore with no significant loss of information.
Right-clicking on a given vertex that has consumed \textsc{cve}
vertices will give the option to reveal the nearest neighbors, including the specific \textsc{cve} entries, indicated in yellow.
Additionally, to indicate how important a given \textsc{cwe} entry in the attack vector space is
the sizing is controlled by how many \textsc{cve} entries it has consumed.
The attack vectors are matched automatically
using another tool that also produces GraphML files~\cite{bakirtzis:2018c},
which implements a search algorithm
to determine which attack vectors could be relevant to the system topology.
The vertex positioning is based on the Frutcherman-Reingold force layout algorithm.
It follows, that the graph layout produces clusters of similar attack vectors 
and, therefore, whole clusters can either be further examined by zooming in
or completely removed (Figure~\ref{fig:dashboard2} \ding{174}).
Finally, the analyst can double-click on the attack vertices
to see all further information in a pop-up window
if needed.
\end{itemize}

All three main panes share actions; that is, most interactive functions
present in the dashboard project to the other panes.
For example, clicking on a vertex in the structure of the specification,
selects the same element in the system topology graph 
and filters the attack vector space only for the attack vectors applicable
to that vertex.

An extra but equally important pane
in the security analyst dashboard is the \textit{bucket} (Figure~\ref{fig:dashboard2} \ding{175}).
The bucket is used as a collection
of attack vectors that the analyst wants
to further investigate over the system topology
or report to the stakeholders.

\begin{figure}[!t]
\centering
\includegraphics[width=1.0\linewidth]{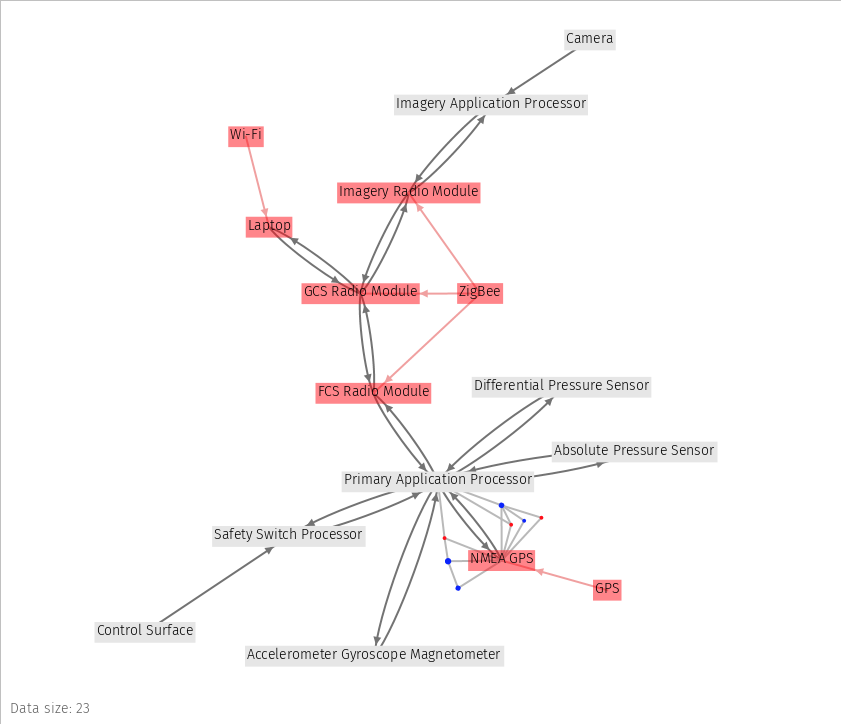}
\caption{Screenshot of the dashboard showing the projection of a selected number of attacks vectors from the bucket. Hovering over the attack vector nodes or zooming further in reveals further information.}
\label{fig:aot}
\end{figure}

Specifically, the bucket contains a table where each row is an attack vector
containing the attack name, a description, and what component(s) it potentially violates.
The rows are color coded the same way as the attack vector graph
to allow for visually identifying between datasets.
The bucket is an essential part of the dashboard as it allows
for the analyst to collect attack vectors based on experience
and, therefore, preemption 
and mitigation have to be considered by the stakeholders.
As an additional feature, the attack vectors in the bucket can be selected
and projected in the system topology; that is, constructs the edges
to the components the attack vector can violate.
This feature can provide further insight;
for example, the analyst might have chosen the attack vector
from filtering based on a specific component but when projecting it
to the system topology finds that it can also violate several other components (Figure~\ref{fig:aot}).

\subsubsection{Data filtering}

When dealing with a large amount
of data---as is the case 
with the attack vector graph, $AV$---it is important 
to implement effective filtering techniques.
To that end, there are several options for filtering data,
the most direct of which is using filter bars.
In the security analyst dashboard there are two filter bars:
one located within the attack vector graph pane (Figure~\ref{fig:paths} \ding{174})
and one located within the bucket pane (Figure~\ref{fig:dashboard2} \ding{175}).
Options to filter include the attack database identification number, name, description,
by the components in which it violates, or all of the above.
Additionally, the search criteria supports regular expressions (RegEx)
to filter attack vectors efficiently.
The filter bar also implements the option to show only the attack vectors visible in the bucket;
that is, all attack vectors that the analyst has deemed important
for further evaluation.

The system topology graph can also be used 
to quickly filter the attack vectors by a specific system topology element vertex
or a set of vertices (Figure~\ref{fig:expoitChains} \ding{172} \& \ding{174}).
This way, the analyst can utilize the three main panes,
by examining the criticality from the systems-theoretic analysis
and then producing evidence to support that the violation has recorded attack vectors.
The analyst can then sort through 
and explore that space normally, since the number 
of attacks to consider is significantly lower.
Finally, the analyst can choose potential attacks
to add to the bucket as reportable artifacts
to system designers and other stakeholders.



\subsubsection{Model \& artifact manipulation}

While the model is predominantly constructed by Systems Engineers,
it is useful for analysts to be able
to manipulate the model's attributes, such that they can constrict the attack vector space.
This is because the attributes
of the system topology model define
the fidelity of the model itself
with respect to the corresponding attack vector space.
Indeed, changing the attributes
of the model is equivalent
to changing the system design itself.
Such minor changes can either show further recorded historic attacks
or, on the other end of the spectrum, reduce the amount of attacks.
Clearly, this mechanism can be used for quick \textit{checkups}
that can then be changed in the model itself 
or discussed with the system designers as viable alternatives
to the current design of the system topology.
An example of such action could be the change
from one specific instance of a sensor to another.
This presumes that a sensor 
with a smaller attack vector space is more secure than one
with a larger attack vector space.

However, being informed about the general attack vector
on the class of the sensor, for example, \textsc{gps},
is also important. 
Attack vectors applicable to a specific \textsc{gps}
that do not match with a different \textsc{gps} implementation
might still be viable.
Model-based security analysis is an iterative learning experience, allowing circumspection about how a system might (mis)behave 
in response to different attack vectors. 
Through this mechanism we attempt to bring \textit{what-ifs}
into security analysis.

An additional action important to an analyst is 
to manipulate the attack vector space directly---through interactive functions like deleting 
or potentially adding vertices to the attack vector space---to 
include information from personal domain-expert experience.
This is because, given the model-based setting, the search results may contain attack vectors 
that do not necessarily apply to the system.
For example, certain attributes might produce weaknesses related to Embedded Java,
but the analyst knows that no software on board the \textsc{cps}
will run Embedded Java. 
In this case, it is useful for the analyst to remove such vertices from the attack vector graph.
For that reason, the dashboard implements the ability 
to delete any number of attack vectors 
by selecting them directly from the graph.

\subsection{Replication}

To allow replication of our results a version 
of the current implementation associated with this paper is hosted on GitHub, including all helper tools to extract and construct the data requirements in use in the security analyst dashboard~\cite{bakirtzis:2018d}.
The dashboard utilizes the Java 8 language framework
to allow for crossplatform operation
of the Graphical User Interface (\textsc{gui}).
To create and render the data as graphs we use GraphStream 1.3~\cite{dutot:2007}.

\section{Example Analysis}
\begin{figure*}[!t]
\centering
\includegraphics[width=1.0\linewidth]{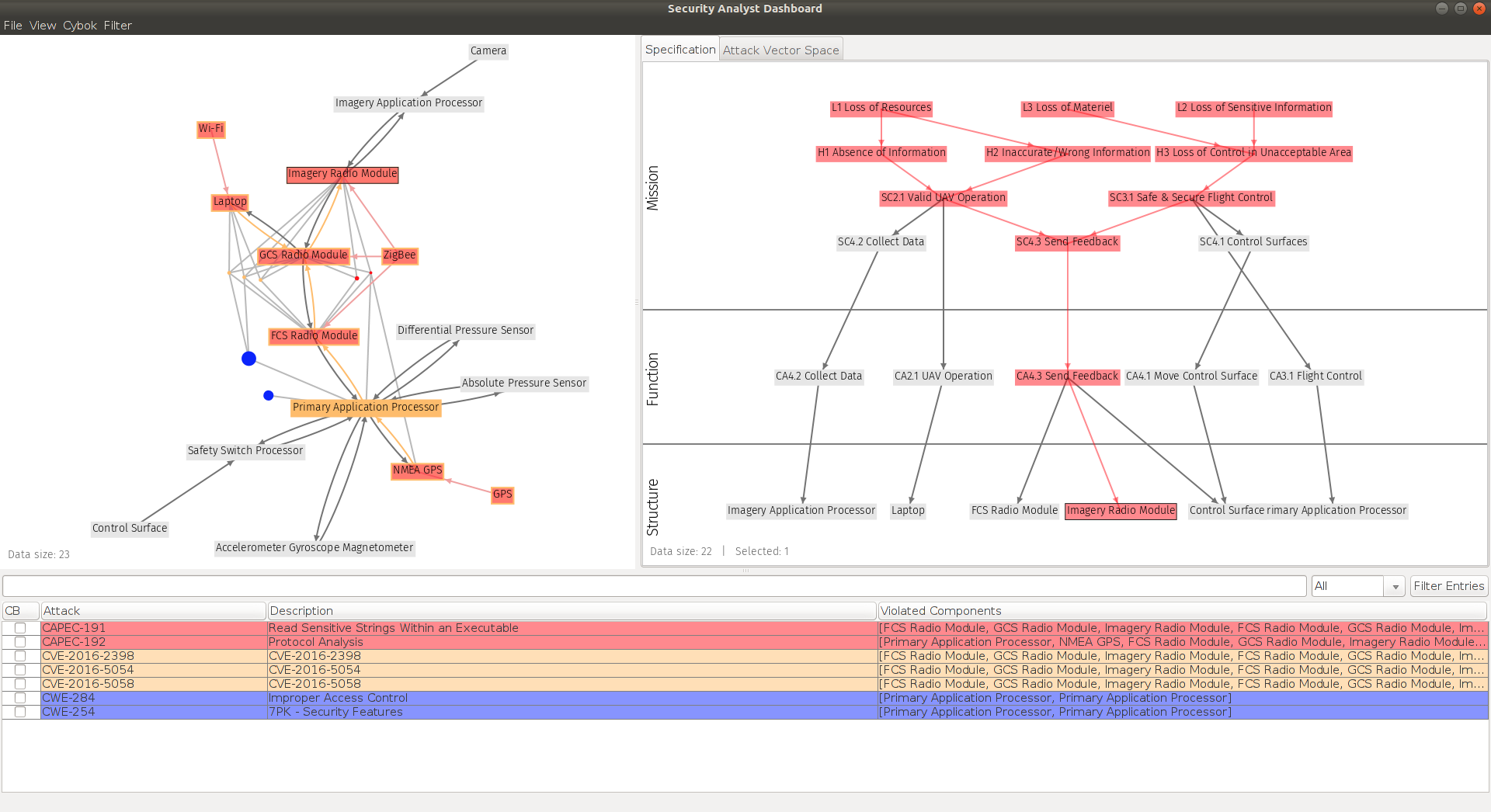}
\caption{Screenshot of the dashboard showing the completed analysis by both a Systems Engineer and security analyst.}
\label{fig:exampleAnalysis}
\end{figure*}

The security analyst dashboard is a natural progression
to model-based security assessment.
Before, we used several other visualization
and data management tools to varying degrees of success.
None of them could by default provide a single view
on the data requirements imposed
by model-based security analysis.
Currently the security analyst dashboard is in use as a research tool
to evaluate different techniques
for safety and security in \textsc{cps}.
A use case study with domain experts would strengthen the results present in this section.
Unfortunately, this is not currently possible because
we have had difficulty finding interested individuals with such narrow domain expertise.
For this reason, this section presents several workflows that
we as researchers found illustrative in sorting through the data necessary
for a holistic assessment
of the system's security posture.

Specifically, to evaluate the security analyst dashboard
we use a model of a \textsc{uas}, its mission specification,
and its associated attack vector space.
Through its attack vector space
we also use the notion
of exploit chain
and attack surface projected 
over the system topology model.

\subsection{Systems-theoretic analysis}
Emma, a Systems Engineer, is tasked with designing a \textsc{cps}
using \textsc{cots} hardware and custom software.
As part of the design process she has constructed a system specification 
and modeled a system topology.
Additionally, she has run the attack vector analysis tool and exported all artifacts
to use in the dashboard (Figure~\ref{fig:dashboard1}).
Emma immediately notices two things.
First, that all radio modules are part
of the attack surface---colored in red---and
also observes that all radio modules share the same violated attribute;
that is, ZigBee (Figure~\ref{fig:dashboard1} \ding{172}).
Second, she notices that the imagery radio module is part
of the system specification, meaning that it is a critical part
of the system.
Before even starting to use the tool Emma is already significantly more informed
than by just looking at the SysML model through the visualizations
of the information produced by the security tool.

Emma clicks on the ``Imagery Radio Module'' vertex on the system specification
and immediately notices that if it is violated it would cause all three unacceptable losses (L1, L2, \& L3)
through the colored paths from the vertex 
to the function; that is, ``CA4.3 Send Feedback,'' all the way up to a subset of safety constraints and
all hazards (H1, H2, \& H3).

Emma decides to report this to other stakeholders, including other security analysts on the team
and Systems Engineers about the possibility of adding defenses if applicable
or considering other resilience mitigation options.

She similarly explores the rest of the interactions
and brings to attention problematic areas based
on her expertise.
If the solution is easy she can change the attributes
of the system model
to further check if the space changes, for example, instead
of using an XBee which requires the ZigBee protocol, she chooses a different radio module,
which does not produce any attack vectors and therefore would not be part of the attack surface
or she might decide to make the system redundant and add multiple different types of radio modules
in the event that other radio modules are equally vulnerable.
By doing so, she builds a strong case for changing the current system topology design
that she can report to the rest of the stakeholders.

\subsection{Attack vector analysis}
Garrett, a security analyst, is not familiar with safety methodologies.
He is, however, intimately familiar with attack vectors; that is,
attack patterns, weaknesses, and vulnerabilities and has extensive experience in embedded \textsc{cps} design
and implementation.
In fact, Garrett has designed
and implemented a \textsc{uas} before,
but had not been aware
of the specific operational goals
of the system.

Through that experience Garrett knows that if the ``Primary Application Processor'' is violated
there would be full mission degradation.
Since Emma has already brought to his attention the potential impact of the radio module,
he combines the two and examines the possible exploit chains
that the security tool has produced for the ``Primary Application Processor''.
By doing so, he quickly finds out that there are several paths from all radio modules,
including the dreaded ``Imagery Radio Module'' 
to the ``Primary Application Processor'' (Figure~\ref{fig:expoitChains} \ding{172}).

Garrett is now significantly more concerned than before about the possibility
that if the radio modules are violated a clear and direct path is possible
both through the vertices and the protocols defining the edges.
He is not so much worried about the ``NMEA GPS,'' from previous experience
and he already has designed hardening techniques
for the Wi-Fi module in the laptop.

Garrett wants to filter down to the attack vectors applicable
to the ``Imagery Radio Module'', he changes to the attack vector space pane
and double-clicks on the ``Imagery Radio Module'' on the system topology pane.
He immediately sees the hundreds of attacks get filtered down
to only the relevant ones.
He zooms in to see the names of the attacks without hovering.
He further searches and double-clicks on \textsc{cwe} entries
to see the neighboors and the specific \textsc{cve} entries.
With a few further searches in the filter bar, he has the worrying attack patterns,
weaknesses, and vulnerabilities cataloged in the bucket.

Additionally, Garrett changes the system topology model slightly
by removing the ZigBee attribute.
By doing so, he notices that the radio modules are not part
of the attack surface but they do still have associated attack vectors.

Garrett can immediately communicate all his findings
to Emma since they both use the same tool by projecting the attacks
in the bucket
to the system topology, a familiar pane to Emma,
who can further read and understand those attacks
by clicking and seeing their description in a pop-up window.

Garrett and Emma conjointly present their findings 
to the rest of the stakeholders using intuitive and effective visualization techniques (Figure~\ref{fig:exampleAnalysis}).
Thus, they provide the stakeholders with defensible, traceable, 
and actionable evidence
for alterations to the current system design.
For instance, the stakeholders might decide
to change the ZigBee radio module completely.
Conversely based on the application 
and taking all information into consideration
they might decide that the ZigBee range is too short
for a practical attack, which then they can capture in their mission-level requirements.

\section{Alternative Approaches}
\label{sec:related}


To our knowledge there is little
to no work merging systems-theoretic analyses
with traditional attack vector analysis
for the visualization of \textsc{cps} designs.

Connective structural 
and mission-oriented information is seen
in \textsc{mitre} CyGraph~\cite{noel:2016},
which does graph-based analytics
and does implement interactivity.
A major deviation from this current work
is that CyGraph is based on more traditional attack graphs~\cite{sheyner:2002},
leverages different notions
of mission impact,
and it mainly targets traditional networked systems.

In traditional attack vector analysis there exist tools
that mine security data
and provide strong search functionality.
One such tool is cve-search~\cite{cve-search},
which includes several more low-level security data repositories, for example, exploit-db~\cite{exploit-db}.
However, cve-search has limited visualization capabilities
and, more significantly, it does not provide interactivity---an important aspect
of an analysis dashboard.

Recently, there has also been interest in the \textsc{mitre} Adversarial Tactics, Techniques \& Common Knowledge (ATT\&CK) framework~\cite{strom:2017,strom:2018}.
This framework allows security analysts
to think in more general terms about consequences
by providing different tables of choices that eliminate consequent choices.
However, the current implementation is not well-suited
for the design and analysis of \textsc{cps},
because there are no external 
or operational goals that match the needs of those systems.

Another approach in the realm 
of model-based security analysis
leverages topic modeling, which immediately relates system models
to relevant \textsc{capec} entries~\cite{adams:2018}.
This approach can compliment the search engine
that produces the data requirements for the dashboard.

In general, a similar rationale to this paper 
for visually constructing 
and presenting models
is shown by Walton et al.~\cite{walton:2015}.
In this work, we target \textsc{cps} directly
and expect the visualizations
to be used from the early stages of the system's lifecycle
and updated often and throughout including deployment and operation,
while Walton et al. target deployment specifically.

The BioFabric~\cite{longabaugh:2012} graph rendering algorithm 
could compliment the current view of the attack vector graph
to deal with the inherent large size and complexity.
This could potentially allow analysts
to better understand how attack vectors relate
to one another, which in turn provides quicker navigation
of applicable attack vectors.

Finally, \textsc{percival}~\cite{angelini:2015} uses similar design choices
(i.e., graph means and projections)
to monitor and analyze computer networks.
However, the domain (traditional information technology (\textsc{it}) systems versus \textsc{cps})
and the representation of exploitation (attack graphs~\cite{shandilya:2014} versus exploit chains) differ.
This is important because the security needs
of a traditional \textsc{it} systems
have significantly different expected consequences.

\section{Concluding Remarks}
In this paper we have proposed a merge between the advances
in systems-theoretic security analysis
and traditional attack vector analysis
for the design and analysis of \textsc{cps}.
We capture this merge in an open-source security analyst dashboard
and show how the complementary views can allow security analysis
throughout the system's lifecycle---from the early-phase and beyond.
Additionally, we show how such a dashboard makes security
an attribute in system design equal
to safety and how, by using a single tool, Systems Engineers
and security analysts can communicate effectively.
By doing this, we promote a proactive approach
to security engineering,
which is increasingly important in the realm of \textsc{cps},
where the consequences 
of security violations lead
to unsafe and uncontrolled behavior.

\section{Acknowledgments}
This research is based upon work supported by the Department of Defense through the Systems Engineering Research Center managed by the Stevens Institute of Technology.
\bibliographystyle{abbrv-doi}
\bibliography{manuscript}

\end{document}